\documentclass[aps,pre,superscriptaddress,twocolumn,groupedaddress,showpacs]{revtex4}
\usepackage{graphicx,epsf}
\begin{document}
\title{Critical dynamics and effective exponents of magnets with extended impurities}
\author{V. Blavats'ka}\email[]{viktoria@icmp.lviv.ua}
\affiliation{Institute for Condensed Matter Physics, National
Academy of Sciences of Ukraine, UA--79011 Lviv, Ukraine}
\author{M. Dudka} \email[]{maxdudka@icmp.lviv.ua}
\affiliation{Institute for Condensed Matter
  Physics, National Academy of Sciences of Ukraine, UA--79011 Lviv,
  Ukraine}
\affiliation{Institut f\"ur Theoretische Physik, Johannes Kepler
Universit\"at Linz, A-4040, Linz, Austria}
 \author{R. Folk}\email[]{folk@tphys.uni-linz.ac.at}
\affiliation{Institut f\"ur Theoretische Physik, Johannes Kepler
Universit\"at Linz, A-4040, Linz, Austria}
\author{Yu. Holovatch}\email[]{hol@icmp.lviv.ua}
\affiliation{Institute for Condensed Matter Physics, National
Academy of Sciences of Ukraine, UA--79011 Lviv, Ukraine}
\affiliation{Institut f\"ur Theoretische Physik, Johannes Kepler
Universit\"at Linz, A-4040, Linz, Austria}
 \affiliation{Ivan Franko National University of Lviv, UA--79005 Lviv, Ukraine}
\date{\today}
\begin{abstract}
 We investigate the asymptotic and effective  static and dynamic critical  behavior
 of ($d=3$)-dimensional  magnets with quenched extended defects, correlated in
 $\varepsilon_d$ dimensions (which can be considered as the
 dimensionality of the defects) and randomly distributed in the
 remaining $d-\varepsilon_d$ dimensions. The field-theoretical
 renormalization group  perturbative expansions being evaluated naively
 do not allow for the reliable numerical data. We apply the Chisholm-Borel
 resummation technique to restore  convergence of the two-loop expansions
 and report the  numerical values of the asymptotic critical exponents
 for the model A dynamics. We discuss different scenarios for static
 and dynamic effective critical behavior and give values for
 corresponding non-universal exponents.
 \end{abstract}
\maketitle

\section{Introduction}
\label{I}

Critical properties of structurally disordered magnets remain a
problem of great interest in condensed matter physics, since
real magnetic crystals are usually non ideal. A simple and natural
case of disorder is implemented via the point-like uncorrelated
quenched non-magnetic impurities and is experimentally realized as
substitutional disorder in uniaxial \cite{Folk03} as well as in
Heisenberg \cite{Pelissetto02,Holovatch02} magnets. Examples are
given by substitute alloys
 ${\rm Mn_{x}Zn_{1-x}F_2}$, ${\rm
Fe_{x}Zn_{1-x}F_2}$ for the uniaxial (Ising) magnets
\cite{Belanger}, and by amorphous magnets ${\rm
Fe_{90+x}Zr_{10-x}}$, ${\rm Fe_{90-y}M_yZr_{10}}$ ( ${\rm
M=Co,\,Mn,\,Ni}$) \cite{ammagn,Perumal01}, transition-metal based
magnetic glasses \cite{magglass,Kellner86} as well as disordered
crystalline materials ${\rm Fe_{100-x}Pt_x}$ \cite{Boxberg94},
${\rm Fe_{70}Ni_{30}}$ \cite{Kellner86} and Eu-chalcogenide solid
solutions \cite{Westerholt} for the Heisenberg magnets. The
question of a great interest arising here is: does the disorder
change critical properties of the systems?  The answer is given by
the famous Harris criterion \cite{Harris74}. It states, that
disorder changes the critical exponents, if the critical exponent
$\alpha_p$ of the pure system is positive: $\alpha_p=2-d\nu_p>0$.
Here, $\nu_p$ is the critical exponent governing the divergence of
correlation length  of the corresponding pure system and $d$ is
the space dimension.

On the base of this inequality, one can estimate the marginal
value $m_c$ for the spin $m$-vector model of magnetic systems,
such that for $m>m_c$ the critical exponents remain unchanged by
point-like defects whereas for $m<m_c$ they cross over to new
values. The present estimates for $m_c$ in three dimensions
definitely imply $m_c<2$: $m_c=1.942\pm0.026$ \cite{Bervillier86}
and $m_c=1.912\pm 0.004$ \cite{Dudka01} and thus only the pure
Ising model ($m=1$) is affected by weak point-like uncorrelated
disorder at criticality.

But in real magnets one encounters non-idealities of structure,
which can not be modelled by simple point-like uncorrelated
impurities. Indeed, magnetic crystals often contain defects of a
more complex structure: linear dislocations, disclinations,
complexes of non-magnetic impurities, embedded in the matrix of
the original crystal \cite{defectbook}. Theoretical studies of
critical behavior of magnets containing such ``extended''
(macroscopic) defects have attracted considerable interest
\cite{Dorogovtsev80,Boyanovsky82,Prudnikov83,Lawrie84,Yamazaki86,
Blavatska02,Blavatska03,Fedorenko04,Weinrib83,lr,mcext,Lee92},
however, a systematic experimental analysis still remains to be
performed.

One of the possible ways to treat defects that extend through the
system being randomly distributed in space but oriented in the
same direction is to consider them as quenched
$\varepsilon_d$-dimensional nonmagnetic impurities of parallel
orientation. This was proposed in the work of Dorogovtsev
\cite{Dorogovtsev80}. The case $\varepsilon_d=0$ is associated
with point-like defects, and extended parallel linear (planar)
defects are described by $\varepsilon_d$ = 1(2). To give an
interpretation to the non-integer values of $\varepsilon_d$, one
may consider patterns of extended defects like aggregation
clusters, and treat $\varepsilon_d$ as the fractal dimension of
these clusters \cite{Yamazaki86}. However, relation of
analytically continued non-integer Euclidean dimension to the
fractal dimension is not straightforward \cite{dimfract}.

\begin{figure}[htbp]
\begin{picture}(220,200)
\epsfxsize=80mm \put(-20,-5){\epsffile[0 0 292 249]{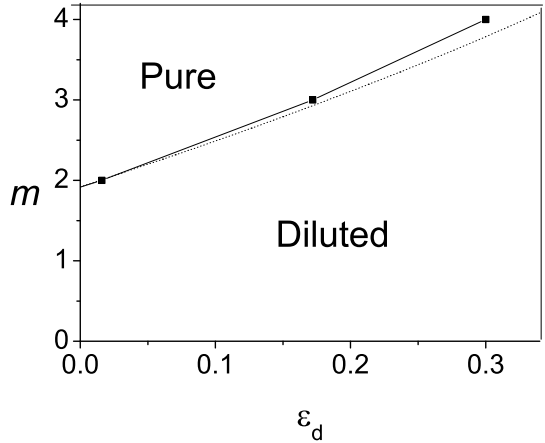}}
\end{picture}
\caption{\label{fig1} The phase diagram of $d=3$-dimensional
$m$-vector magnets in the presence of $\varepsilon_d$-dimensional
impurities. Marginal value $m_c(\varepsilon_d)$ separates regions
of different universality classes. Disorder is relevant in the
$\varepsilon_d$-$m$ plane below the $m_c$ curve. Filled squares:
results of calculations, based on six-loop renormalization group
data \cite{Guida98} for the correlation length critical exponent,
dotted line: estimate based on resummed $\varepsilon,
\varepsilon_d$-expansion for $m_c$, Eq. (\ref{mcryt}). Only the
physical region $\varepsilon_d>0$ is shown. See Section \ref{IV}
for a whole description.}
\end{figure}

It was shown \cite{Boyanovsky82}, that presence of extended
impurities leads to a generalized Harris criterion, namely, in
this case disorder alters the critical behavior of the pure
system, if $ \varepsilon_d>d-2/\nu_{p}$. If one considers the
point-like disorder with $\varepsilon_d=0$, the generalized Harris
criterion turns to ordinary one cited above. Again, this
inequality  defines for each value of $\varepsilon_d$ the critical
value $m_c$, below which the extended disorder alters the
universality class. But now the disorder induced critical behavior
of $d=3$ systems holds not only for the Ising magnets ($m=1$) but
for $m>1$ as well, as shown in Fig. \ref{fig1} (region denoted as
``Diluted" in the figure). Therefore, the class of magnets where
the new critical behavior may be found is not restricted to the
uniaxial ones ($m=1$), but include easy-plane ($m=2$) and
Heisenberg ($m=3$) systems. Thus predicted phenomena of a new
universal behavior may be experimentally checked for a wider class
of magnets.

Another interesting feature of systems with parallel extended
defects is that due to the spatial anisotropy they are described
by  two correlation lengths, one perpendicular, $\xi_{\perp}$, and
one parallel,  $\xi_{||}$, to the extended impurities direction
\cite{Dorogovtsev80}. As the critical temperature $T_c$ is
approached, their divergences are characterized by corresponding
critical exponents $\nu_{\perp}$, $\nu_{||}$:
\begin {equation}
\label{xi} \xi_{\perp}\sim |t|^{-\nu_{\perp}},\phantom{5555555}
\xi_{||}\sim |t|^{-\nu_{||}},
\end{equation}
where $t$ is the reduced distance to the critical temperature
$t=(T-T_c)/T_c$. Anisotropic scaling holds also for the spin-spin
pair correlation function at $T_c$ and is governed by exponents
$\eta_{\perp}$ and $\eta_{||}$. Whereas the magnetic
susceptibility is isotropic, as far as all order parameter
components  interact with defects in a similar way. However, the
dynamic critical behavior is modified, two times of relaxation
in directions perpendicular and parallel to the extended impurities
$\tau_{\perp}$ and $\tau_{||}$  behave correspondingly as:
\begin {equation}
\label{tau} \tau_{\perp}\sim
\xi_{\perp}^{z_{\perp}},\phantom{5555555} \tau_{||}\sim
\xi_{||}^{z_{||}},
\end{equation}
 with dynamical exponents $z_{\perp}$ and $z_{||}$.

Therefore, magnets with quenched extended defects of parallel
orientation constitute a large class of systems with a bulk of
unusual phenomena worth to be analyzed. Their asymptotic critical
behavior was  a subject of the field-theoretical renormalization
group (RG) analyses
\cite{Dorogovtsev80,Boyanovsky82,Prudnikov83,Blavatska02,Blavatska03}.
In particular, a double expansion in both $\varepsilon=4-d$,
$\varepsilon_d$ was suggested and RG functions were calculated
\cite{Dorogovtsev80} to order $\varepsilon$, $\varepsilon_d$;
qualitatively, the crossover to a new universality class in the
presence of extended defects was supported. These calculations
were extended to the second order in Refs.
\cite{Boyanovsky82,Lawrie84}. However, these divergent RG
expansion did not give a reliable numerical estimate for the
critical exponents - the goal highly desirable both for
experimental and simulational purposes. Numerical estimates for
the exponents describing {\em static} critical behavior of $d=3$
magnets with parallel extended impurities were obtained only
recently \cite{Blavatska02,Blavatska03} by applying special
resummation technique to the two-loop RG functions of Refs.
\cite{Boyanovsky82,Lawrie84}. Currently, no estimates of similar
accuracy exist for the {\em dynamic} exponents describing critical
slowing down in the magnets under consideration. The  theoretical
investigation of critical dynamics of system with parallel
extended defects was performed in the one-loop approximation
\cite{Prudnikov83} for the simple models with non-conserved order
parameter (model A) and conserved order parameter (model B). Then
the RG study of model A critical dynamics was extended to the
two-loop order \cite{Lawrie84}.  But again, the convergence
properties of the series obtained did not allow numerical
evaluations. Recently, the short-time critical dynamics of the
model A was considered in Ref. \cite{Fedorenko04}.

The goal of this paper is to apply the state-of-the-art analysis
of the divergent RG perturbation theory series to get numerical
estimates of the exponents describing model A dynamics of $d=3$
magnets with extended impurities for a wide region of impurity
dimension $\varepsilon_d$ and for different values of $m$.
Complementing existing estimates of the static exponents
\cite{Blavatska02,Blavatska03}, our results will give
comprehensive values for the critical exponents of systems with
extended impurities and, in this way, will facilitate their
experimental analysis. Moreover, since both in experiments and in
Monte Carlo (MC) simulations often the asymptotic region is not reached and
effective exponents are observed \cite{effective}, we will
calculate these as well. By these estimates we will predict the
possible scenarios of approaching the critical point.
These are important to perform corresponding experimental checks,
as it was shown recently by theoretical \cite{Dudka03} and
experimental \cite{expereff} studies of the $d=3$ disordered
Heisenberg magnets with point-like defects.

The setup of the paper is as follows. In the next section we
present the model; in section \ref{III}  the renormalization
procedure is discussed. In section \ref{IV} we apply the
resummation techniques to analyze the renormalization group
functions in two-loop approximation and present the quantitative
estimates for the asymptotic dynamical critical exponents. Section
\ref{V} gives description of  possible scenarios for the effective
critical behavior and section \ref{VI} concludes our study.

\section{The model}
\label{II}

The starting point is the effective Hamiltonian of the model of an
$m$-vector magnet with $\varepsilon_d$-dimensional defects,
extending throughout the system along the coordinate directions
symbolized as $x_{||}$ and randomly distributed in perpendicular
directions $x_{\perp}$ \cite{Boyanovsky82,Lawrie84}:
\begin{eqnarray}
{\cal H}&=&\int d^dx\big[\frac{1}{2}(\mu_0^2+V(x))
\vec{\phi}^2(x)+ (\nabla_{\perp} \vec{\phi}(x))^2 \nonumber\\ &&
+a_0({\nabla}_{||}
\vec{{\phi}}(x))^2+\frac{u_0}{4!}(\vec{\phi}^2(x))^2\big].
\end{eqnarray}
Here, $\vec{\phi}$ is an $m$-component vector field:
$\vec{\phi}=\{ \phi^{1}\cdots\phi^{m}\}$, $\mu_0$ and $u_0$ are
the bare mass and the coupling, $a_0$ is the bare anisotropy
constant, $\nabla_{||}$ and $\nabla_{\perp}$ denote
differentiation in the coordinates $x_{||}$  and $x_{\perp}$ and
the impurity potential $V(x)$ is introduced. The probability
distribution of defects has zero mean and variance given by:
\begin{eqnarray}
\langle\langle V(x)V(y)\rangle\rangle&
=&-v_0\delta^{d-\varepsilon_d}(x_{\perp}-y_{\perp}). \,\,\,\,\,\,
\label{corr}
\end{eqnarray}
Here, $\langle\langle...\rangle\rangle$ denotes the averaging over
the distribution of defects, (-$v_0$) is a positive coupling
constant, which is proportional to the concentration of
impurities. The constant $a_0$ parameterizes the anisotropy,
arising in the system due to the presence of extended defects.

We consider critical dynamics of the model (\ref{model}) for the
case of a non-conserved order parameter. For this case the dynamics
can be expressed in the Langevin equation form \cite{Hohenberg77}:
\begin{equation}
\frac{\partial \phi^{i}(x,t)}{\partial t}=-\lambda_0
\frac{\partial {\cal H}}{\partial
\phi^{i}(x,t)}+\eta^i(x,t),\phantom{555}i=1...m,
\end{equation}
where $\lambda_0$ is the Onsager kinetic coefficient and
$\eta^i(x,t)$ is the Gaussian random-noise source with zero mean
and correlation:
\begin{equation}
\langle \eta^i(x,t)\eta^j(x',t')\rangle =2\lambda_0
\delta(x-x')\delta(t-t')\delta_{ij}.
\end{equation}
The brackets $\langle \dots\rangle$ mean an average with respect
to the thermal noise.

Within the field theory approach it is convenient to use the
Bausch-Janssen-Wagner formulation \cite{Bauch76} which is given by
the Lagrangian:
\begin{equation}
{\cal L}[\tilde\phi,\phi]= \int d^d x\, dt\,\sum_i
\tilde\phi^i\left[\frac{\partial\phi^i}{\partial
t}+\lambda_0\frac{\delta {\cal H}}{\delta
\phi^i}-\lambda_0\tilde\phi^i\right].
\end{equation}
Here, $\tilde\phi^i$ are components of an auxiliary response field
introduced to average over the thermal noise. Then correlation and
response functions are computed with the help of a weight function
$W\sim e^{-{\cal L}[\tilde\phi,\phi]}$.

It is known that studying dynamical properties of disordered
systems averaging over random impurities can be applied directly
to dynamical weight function $W$. As it is established in Ref.
\cite{DeDominicis78}, the configurational averaging can be
performed avoiding replica trick \cite{Emery75}. However it leads
to the same perturbative expansions for the RG functions as those
obtained with replica formalism. Therefore both approaches giving
equivalent results are possible for dynamics. However the above
alternative does not exist when the static critical behavior is
analyzed (see e.g. \cite{Folk03,Pelissetto02}). As far as both
static and dynamic criticality is addressed in this paper,  we use
the replica trick, representing the logarithm of a weight function
in the following form:
\begin{equation}
\ln W=\lim_{n\to 0}\frac{\langle\langle W^n-1 \rangle\rangle}{n}.
\end{equation}
Finally it leads to study of properties of replicated Lagrangian
${\cal L}[\tilde\phi,\phi]$ \cite{note}:
\begin{widetext}
\begin{eqnarray}
{\cal L}[\tilde\phi,\phi]&=&\sum_{\alpha}\Bigg\{ \int d^d x
dt\,\,\sum_i \tilde\phi^i_{\alpha}\left[
\frac{\partial\phi^i_{\alpha}}{\partial
t}+\lambda_0(\mu^2_0-\nabla_{\bot}^2-a_0\nabla_{\|}^2)
\phi^i_{\alpha}-\lambda_0\tilde\phi^i_{\alpha}
+\sum_{j}\lambda_0\frac{u_0}{3!}
\phi^i_{\alpha}\phi^j_{\alpha}\phi^j_{\alpha}\right]\nonumber\\
&&+\sum_{i,j,\beta}\lambda^2_0\frac{v_0}{2}\int d^d x\,d^d y\,
dt\, dt' \delta(x_{\bot}-y_{\bot})
\phi^i_{\alpha}(x,t)\phi^i_{\alpha}(x,t)\phi^j_{\beta}(y,t')\phi^j_{\beta}(y,t')\Bigg\}\label{model},
\end{eqnarray}
\end{widetext}
here, the summation over Greek indices spans from 1 to $n$ denoting the
different replicas and the
Latin indices go from 1 to $m$ denoting the components of the
order parameter.  To study behavior of this
model (\ref{model}) in the vicinity of the critical point we apply
the minimal subtraction scheme within field-theoretical RG. The
description of this approach is given in the next section.

\section{RG study}
\label{III}

The description of  the long-distance properties of the model
(\ref{model}) near the second order phase transition point is
performed using the field-theoretical RG method \cite{rgbooks}.
Let us present the renormalization algorithm, developed for
Lagrangian field theory. It is well known, that such a theory
encounters with ultraviolet divergences, the removal of which is
achieved within an appropriate renormalization procedure by a
controlled rearangement of the perturbation theory series. The
change of bare couplings $u_0,v_0$ and of the anisotropy constant
$a_0$ under renormalization is described by the RG functions:
\begin{eqnarray}\label{13a}
\beta_u(u,v)&=&\frac{\partial u}{\partial \ln \kappa}|_0, \,\,
\beta_v(u,v)=\frac{\partial v}{\partial \ln \kappa}|_0,\\
\zeta_{a}(u,v)&=&\frac{\partial \ln a}{\partial \ln \kappa}|_0.
\label{13c}
\end{eqnarray}
Here, $u, v, a$ are the renormalized couplings and anisotropy
constant, respectively, $\kappa$ is the rescaling parameter, the
notation $|_0$ indicates differentiation at fixed bare parameters.
The bare fields $\phi$, $\tilde{\phi}$, the mass $\mu_0$ and the
Onsager kinetic coefficient $\lambda_0$ are related to the
renormalized ones $\varphi$, $\tilde{\varphi}$ $\mu$ and $\lambda$
by:
\begin{eqnarray} \nonumber
\phi&=&Z_{\varphi}^{1/2}\varphi, \hspace{2em}
\tilde{\phi}=Z_{\tilde{\varphi}}^{1/2}\tilde{\varphi},
\\ \nonumber
\mu_0^2&=&Z_{\mu^2}\mu^2, \hspace{2em}
\lambda_0^{-1}=Z_{\lambda}\lambda^{-1},
\end{eqnarray}
where $Z$-factors are dimensionless functions of renormalized
parameters $m,u,v,a$. Their flows are defined by corresponding RG
functions:
\begin{eqnarray} \label{14a}
 \gamma_{\phi}(u,v)&=&\frac{\partial \ln Z_{\varphi}}{\partial \ln
 \kappa}|_0, \\
\label{14aa}
 \gamma_{\tilde\phi}(u,v)&=&\frac{\partial \ln Z_{\tilde\varphi}}{\partial \ln
 \kappa}|_0, \\  \label{14b}
 {\bar \gamma}_{\phi^2}(u,v)&=&\frac{\partial \ln
 Z_{\mu^2}^{-1} }{\partial \ln \kappa }|_0-\gamma_{\phi}, \\
  \label{14c}
 \zeta(u,v)&=&\frac{\partial \ln Z_{\lambda}}{\partial \ln
\kappa}|_0.
\end{eqnarray}
Functions $\zeta$ and $\gamma_{\tilde\phi}$ are connected by the
relation: $\zeta=(\gamma_{\phi}-\gamma_{\tilde\phi})/2$.

The fixed points (FPs) $u^*,v^*$ of the RG transformation are
defined as common zeroes of the $\beta$-functions:
\begin{equation} \beta_u(u^*,v^*)=0, \,\, \beta_{v}(u^*,v^*)=0.\label{fp}
\end{equation}
 A FP is
stable, if the eigenvalues of the stability matrix, defined as: $
B_{ij}={\partial \beta_{u_i}(u^*,v^*)}/{\partial
\beta_{u_j}(u^*,v^*)},\,u_i=\{u,v\} $ have positive real parts. In
the case when the stable FP is physically accessible, i.e. can be
reached starting from the initial values of the renormalized
couplings (in our case, $u>0$, $v<0$), corresponds to the critical
point of the system. The RG functions (\ref{14a})-(\ref{14c})
taken at this point give the critical exponents of magnetic
susceptibility, correlation length and relaxation time:
\begin{eqnarray}
\gamma^{-1}&=& 1-\frac{ {\bar \gamma}_{\phi^2}(u^*,v^*) }
{2-\gamma_{\phi}(u^*,v^*)},\nonumber\\
 \nu_{\perp}^{-1}&=&2- {\bar
\gamma}_{\phi^2}(u^*,v^*)-\gamma_{\phi}(u^*,v^*),\label{exp}\\
z_{\perp}&=&2+\zeta(u^*,v^*).\label{zperp}
\end{eqnarray}
Note, that the following relations between the exponents
describing the parallel and perpendicular correlation length and
relaxation times hold \cite{Dorogovtsev80}:
\begin{eqnarray}
\nu_{||}&=&\nu_{\perp} (1-\frac{\zeta_{a}}{2})\nonumber,\\
z_{||}&=&z_{\perp}/(1-\frac{\zeta_{a}}{2})\label{zpar}.
\end{eqnarray}

To obtain the quantitative characteristics of the dynamical
critical behavior of magnetic systems with extended impurities, we
turn our attention to the RG functions derived in Ref.
\cite{Lawrie84} in two-loop approximation:
\begin{eqnarray}
{\beta_u}/{u}&=&-\varepsilon+\frac{(m+8)}{6}u+2v-
\frac{(3m+14)}{12}{u}^{2} \nonumber\\ &-& \frac{1}{12}uv\left[
\frac{2}{3}(11m+58)+(m-4)\frac{\varepsilon_d}{3(\varepsilon+
\varepsilon_d)}\right] \nonumber\\ &-& v^2\frac{1}{144}\left[
328+32\frac{\varepsilon_d}{\varepsilon+\varepsilon_d}\right],
\label{betau}\\
{\beta_{v}}/{v}&=&-\varepsilon-{\varepsilon_d}+\frac{4}{3}v+
\frac{m+2}{3}u-\frac{7}{6} v^2\nonumber\\ &-&v
u\frac{m+2}{18}\left[11-\frac{\varepsilon_d}{\varepsilon+
\varepsilon_d}\right]-\frac{5}{12}\frac{m+2}{3}u^2,
 \label{betav}\\
\gamma_{\phi}&=& \frac{1}{36}v^2+\frac{m+2}{36}v u
+\frac{m+2}{72}u^2,
 \label{gammaf}\\
{\bar
\gamma}_{\phi^2}&=&u\frac{m+2}{6}+\frac{1}{3}v-\frac{m+2}{6}u^2-24v^2
\nonumber\\ &-&\frac{m+2}{24}v u
\left[6-\frac{\varepsilon_d}{\varepsilon+ \varepsilon_d}
\right],\label{gammaf2}\\
\zeta_{a}&=&\frac{1}{3}v-\frac{5}{36}v^2-\frac{(m+2)}{36}v\,u,
\label{zetaa}\\ \nonumber \zeta&=&-\frac{1}{3}v+
\frac{(m+2)(6\ln\frac{4}{3}-1)}{72}u^{2}+ \\ &&
 \frac{(m+2)}{36}v u+\frac
{5}{36}{v}^{2}.\label{zeta}
\end{eqnarray}
Here, $\varepsilon=4-d$ and the replica limit $n=0$ has been
assumed. Putting in Eqs. (\ref{betau})-(\ref{gammaf2}),
(\ref{zeta}) $\varepsilon_d=0$ one regains the RG functions of the
$m$-vector magnet with point-like defects
\cite{Janssen95,Kleinert95,Folk00}.

The description of the critical behavior of magnets with extended
impurities carried out in Refs.
\cite{Dorogovtsev80,Boyanovsky82,Lawrie84,Yamazaki86} was based on
the double expansion in two parameters $\varepsilon$,
$\varepsilon_d$, considering both to be of the same order of
magnitude. Being quite clear technically, such a statement causes
however certain cautions. Indeed, taking the $\varepsilon$ as a
small parameter is due to the fact that it deviates from the upper
critical dimension, where the $\phi^4$ theory becomes
asymptotically free \cite{rgbooks}, whereas $\varepsilon_d$ is the
dimension of defect itself and obviously has a different physical
origin. Therefore, it is desirable to search for an alternative
way to analyze the RG functions (\ref{betau})-(\ref{zeta}).
Fortunately, such an alternative exists and it is exploited in the
field-theoretical approach to critical phenomena \cite{Schloms}.
Namely, it consists in the analysis of minimal subtraction RG
functions directly at dimension of space of interest, fixing
$\varepsilon$ and solving the FP equations numerically
\cite{Schloms} (the so-called fixed dimension scheme).  It was
proposed \cite{Blavatska02,Blavatska03} to extend the approach of
direct evaluation to the RG functions of the present model, i.e.
to treat them directly at $d=3$ $(\varepsilon=1)$ for different
fixed values of the (non-integer) defect dimensionality
$\varepsilon_d$. In our calculations (see Section \ref{IV}), we
will make use of both possibilities, exploiting $\varepsilon$,
$\varepsilon_d$ expansion as well as working within the fixed
dimension scheme.

In the field-theoretical RG approach the series expansions of the
RG functions in powers of the coupling appear to be divergent;
moreover, they are characterized by a factorial growth of the
coefficients implying a zero radius of convergence \cite{rgbooks}.
To take into account the higher order contributions, the
application of special tools of resummation is required
\cite{Hardy48}. In our previous paper \cite{Blavatska03},
analyzing static critical behavior we applied the Chisholm-Borel
resummation technique. Here, the Borel image of the initial
function is extrapolated by a rational Chisholm \cite{Chisholm73}
approximant $[K/L](x,y)$. One constructs this ratio of two
polynomials of order $K$ and $L$ such that its truncated Taylor
expansion is equal to that of the Borel image of the initial
function. The resummed function is then calculated by an inverse
Borel transform of this approximant. The details can be found in
\cite{IJMP}.

\section {The results}
\label{IV}

The critical behavior of the model (\ref{model}) is influenced by
the presence of the following FPs: the Gaussian FP {\bf G} ($u^*=v^*=0$), the
$O(m)$-symmetric FP of a pure magnet {\bf P} ($u^*\neq0, v^*=0$)
and the random FP {\bf R} ($u^*\neq 0, v^*\neq 0$) that governs
disorder-induced critical behavior. A polymer FP with $u^*=0,
v^*>0$ is not reachable from the initial values of couplings and
therefore is out of interest for our study. Depending on the
values of global parameters $d,\varepsilon_d, m$ one of the above
FPs is stable and in the asymptotic limit governs the criticality.
However, as we will show in the section \ref{V} an approach to the
asymptotic limit and hence the effective critical behavior is
influenced by all the FPs.

For the $d=3$ magnets considered here, it is the crossover between
FPs {\bf P} and {\bf R} that corresponds to the change of the
universal properties of an $m$-vector magnet upon dilution by
$\varepsilon_d$-dimensional defects. According to the generalized
Harris criterion, this crossover occurs at certain marginal value
$m_c(\varepsilon_d)$. At $m>m_c$, the FP {\bf P} is stable,
indicating no change in critical behavior, whereas for $m<m_c$ one
finds the stability of the FP {\bf R}, displaying the fact of
relevance of disorder. From the generalized Harris criterion one
can obtain the marginal value $\varepsilon_d^{{\rm marg}}$ as
function of $m$. Using the six-loop results \cite{Guida98} for the
correlation length critical exponent $\nu_p(m)$ of the pure
$m$-vector magnet, one obtains: $\varepsilon_d^{{{\rm marg}}}(m=1)
=-0.173$; $\varepsilon_d^{{{\rm marg}}}(m=2)=0.016$;
$\varepsilon_d^{{{\rm marg}}}(m=3) =0.172$; $\varepsilon_d^{{{\rm
marg}}}(m=4) = 0.300$. These estimates are shown in Fig.
\ref{fig1} by filled squares connected by a solid line. The figure
may serve as a phase diagram of an $m$-vector magnet with extended
impurities: the new critical behavior is expected in the region of
$\varepsilon_d - m$ plane denoted as ``Diluted". In the region
denoted ``Pure" the asymptotic critical behavior does not indicate
presence of defects, however the effective critical behavior does.
Below, we will perform an analysis of the critical dynamics for
both regions of the phase diagram.

\subsection{Fixed $d=3$ scheme}\label{IVA}

We start from the RG $\beta$-functions (\ref{betau}),
(\ref{betav}): fixing the value $\varepsilon=1$ (i.e. $d=3$) and
treating $\varepsilon_d$ as a varying parameter, we look for the
common zeros of the resummed functions $\beta_u$ and $\beta_v$.
 The numerical values of the stable FP coordinates
obtained for the three-dimensional $m$-component magnets with
$m=1,2,3,4$ can be found in our previous paper \cite{Blavatska03}.

To calculate the values of the critical exponents $z_{\perp}$ we
substitute Eq. (\ref{zeta}) for $\zeta$ in Eq. (\ref{zperp}),
apply the resummation procedure for the resulting series, and,
finally, estimate them at the stable FP. Note, that because the
function $\zeta$ is not symmetric in variables $u$ and $v$
(namely, it does not contain term linear in $u$) the Chisholm
approximant chosen for its analytic continuation differs from
those chosen for the $\beta$-functions.  The value of critical
exponent $z_{||}$ is obtained using the relation (\ref{zpar}).

\begin{table}[h!]
\begin{center}
\caption{\label{tab1} The values of dynamical critical exponents
of three-dimensional $m$-component magnets at different fixed
values of extended defect dimension $\varepsilon_d$.}
\label{expon}
\begin{tabular}{c c  c  c  c c  c c  c}
\hline\hline  &   &  $m=1$  &   & $m=2$ &  &$m=3$&  &
$m=4$\\\hline $\varepsilon_d$  & $z_{\perp}$  &  $z_{||} $  &
$z_{\perp}$ & $z_{||} $  &  $z_{\perp}$  & $z_{||} $ & $z_{\perp}$
& $z_{||} $
\\
0 & 2.172 & - &2.065 & -  & 2.062 & - &2.058 & -
 \\
 0.1 & 2.248 & 2.094 &2.079  &2.068 &2.062 & 2.062 &2.058& 2.058
 \\
 0.2 & 2.302 & 2.111 &2.135 & 2.073 & 2.062& 2.062 &2.058& 2.058
\\
0.3 & 2.340 & 2.127 &2.183 & 2.074 & 2.095 & 2.063& 2.058 & 2.058
\\
0.4 & 2.367 & 2.142 &2.223 & 2.081 &2.133 & 2.063 &2.083 & 2.061
\\
 0.5 &2.386 & 2.156 & 2.256 & 2.087 & 2.166 & 2.064&2.111 & 2.062\\
 0.6,& 2.399 & 2.169 &2.284 & 2.092 &2.195 & 2.067 &2.138 & 2.065
\\
0.7 & 2.408 & 2.181 &2.307 & 2.096 &2.219 & 2.069 &2.162 & 2.067
\\ 0.8 & 2.413 & 2.194 &2.326 & 2.100 &2.241 & 2.070
 &2.184 & 2.069 \\
 0.9 & 2.416 & 2.206
&2.342 & 2.104 &2.260 & 2.072 &2.203 & 2.070
\\ 1.0 & 2.418 &
2.217 &2.356 & 2.107 & 2.276 & 2.073&2.219 & 2.071\\ 1.1 & 2.418,&
2.228  & 2.368 & 2.109 &2.290 & 2.073 & 2.233 & 2.071
\\ \hline\hline
\end{tabular}\end{center}
\end{table}

In Table \ref{tab1} we give the obtained results for critical
exponents $z_{||}$, $z_{\perp}$ of three-dimensional $m$-component
magnets.  The case $\varepsilon_d=0$ corresponds to point-like
defects. As it was already noted in the Introduction, for this
case only the Ising model ($m=1$) is influenced by disorder, and
thus for systems with $m>1$ the values of the critical exponents
are not altered by disorder. Here, one can compare our result with
the dynamic exponent of the pure and diluted Ising models, as
shown in Tables  \ref{tab2}, \ref{tab3}.
 When $\varepsilon_d$ increases,
for $m=2,3,4$ the critical exponents remain constant and equal to
the corresponding exponents of the pure model, until $m$ becomes
 $m_c$ for given $\varepsilon_d$ and  for $m>m_c$ the
values of exponents start to increase because they take their new
values belonging to the disordered universality class.

\begin{table}[h!]
\caption{\label{tab2} The value of dynamical critical exponent $z$
for pure three-dimensional Ising model. MC: Monte Carlo
simulations; exp: experimental estimates for ${\rm FeF_2}$.
Theoretical estimates, $\sim\varepsilon^2(\varepsilon^3)$: direct
substitution $\varepsilon=1$ into
 $\sim\varepsilon^2(\varepsilon^3)$-expansions; $\sim\varepsilon^3$, res: resummation
of $\varepsilon^3$ expansion; 4 loops, res: resummation of
four-loop massive 3d RG functions.}
\begin{center}
\begin{tabular}{ccc}
\hline \hline &Method&Result\\ \hline
 Ref. \cite{Wansleben91}&MC &$2.04\pm0.03$\\
 Ref. \cite{Ito93}&MC &$2.06\pm0.02$\\
 Ref. \cite{Kikuchi93}&MC &$2.03\pm0.01$\\
 Ref. \cite{Grassberger95}&MC &$2.032\pm0.004$\\
 Ref.  \cite{Ito00}&MC & $2.055\pm0.003$ \\
 Ref.  \cite{Belanger88}&exp & $2.1\pm 0.1$ \\
   Ref. \cite{Lawrie84}&$\sim \varepsilon^2$ & 2.014\\
 Eq. (\ref{3la}) & $\sim \varepsilon^3$&2.024\\
 Eq. (\ref{3la}) & $\sim \varepsilon^3$, res & 2.012\\
 Ref. \cite{Prudnikov97}&4 loops, res&2.017\\
 \hline\hline
 \end{tabular}
 \end{center}
\end{table}

\begin{table}[h!]
\caption{\label{tab3} The value of dynamical critical exponent $z$
for three-dimensional Ising model with point-like uncorrelated
defects. MC: Monte Carlo simulations; exp: experimental estimates
for ${\rm Fe_{0.9}Zn_{0.1}F_2}$.
  Theoretical estimates,
$\sim\varepsilon^{1/2}$: direct substitution $\varepsilon=1$ into
 $\varepsilon^{1/2}$-expansions; 2 loops, res: resummation of two-loop
 massive 3d RG functions; $\sim\varepsilon^{1/2}(\varepsilon^{3/2})$ res: substitution
of the two (three)-loop resummed FP coordinates into the two-loop
expansions for $z$. }
\begin{center}
\begin{tabular}{ccc}
\hline \hline &Method&Result\\ \hline
 Ref. \cite{Prudnikov92}& MC & $2.13\pm 0.05$ \\
 Ref. \cite{Heuer93} &MC  &$2.4\pm0.1$\\
 Ref. \cite{Parisi99} &MC & $2.62\pm 0.07$\\
 Ref. \cite{Rosov92} &exp & $2.18\pm 0.10$\\
 Ref. \cite{Grinstein77} & $\sim \varepsilon^{1/2}$ &2.336\\
 Ref. \cite{Prudnikov92}& 2 loops, res & 2.237 \\
 Ref.  \cite{Janssen95}&$\sim \varepsilon^{1/2}$, res &2.023\\
 Ref. \cite{Oerding95}& $\sim \varepsilon^{3/2}$, res &2.180\\
 \hline\hline
 \end{tabular}
 \end{center}
\end{table}

An interesting feature of the data shown in Table \ref{tab1} is that
for each fixed $\varepsilon_d$ the relation  $ z_{||}<z_{\perp}$
holds. One can give the following  physical interpretation to this
fact. As it has been noted for isotropic systems
\cite{Hohenberg77}, the dynamical critical exponent $z$ is
proportional to the pair correlation function critical exponents
$\eta$:
 \begin{equation} \label{eta}
 z=2+c\eta.
 \end{equation}
For the systems we consider here the critical exponents
$\eta_{\perp}$ and $\eta_{||}$, that  characterize the behavior of
the pair spin-spin  correlation  function in the directions,
perpendicular and parallel to the extended defects, are
distinguished  \cite{Dorogovtsev80}. As far as the extended
defects cut interacting bonds of spins perpendicular to the
extended-defect direction,  in the parallel direction the
fluctuations are stronger and therefore $\eta_{||}<\eta_{\perp}$
and thus, by Eq. (\ref{eta}) with $z_{||}$, $z_{\perp}$ we obtain
the confirmation to our results. However, in the anisotropic case the coefficient
$c$ in Eq.~(\ref{eta}) is also direction dependent. Thus the above argumentation is
rather of qualitative nature to give a hint to physical interpretation
of the observed relation $ z_{||}<z_{\perp}$.

In our previous papers \cite{Blavatska02,Blavatska03} we have also
touched the question of existence of an upper marginal value for
the defect dimensionality $\varepsilon_d$. In our analysis, we
observe the disappearance of a stable reachable FP for
$\varepsilon_d$ slightly above $1$. The following physical
interpretation was proposed: extended defects of large dimension
(e.g. parallel planar defects with $\varepsilon_d=2$), may divide
the system into non-interacting regions and thus inhibit
ferromagnetic order. Recently in Refs. \cite{Fendler05} the Ising
magnet with spin interaction bonds, correlated in 2 dimensions and
randomly distributed in the perpendicular direction, was studied.
Although such a system differs from those considered here, both
models possesses a number of common features. In particular, the
smearing of a phase transition due to the presence of planar
defects was predicted and explained by the existence - although
rare - of infinite spatial regions, which are free of defects and
therefore may be locally in the ordered phase.

 To confirm the obtained results, we have also tried to
apply the Pad\'e-Borel resummation to estimate the values for
$z_{\perp}$, $z_{||}$. In this procedure, one rewrites the
two-variable series (\ref{betau}), (\ref{betav}) in terms of a
resolvent series \cite{Watson74} of one variable and then applies
a Pad\'e approximant for its analytic continuation. We do not
present the results obtained, they give rather close numbers to
those in Table \ref{tab3} and reproduce the same behavior.

\subsection{$\varepsilon,\varepsilon_d$-expansion}\label{IVB}

Another possibility to obtain estimates for dynamic critical
exponents of the $m$-vector magnet with extended defects is to
resum the double $\varepsilon,\varepsilon_d$ expansions obtained
in the two-loop approximation in Ref. \cite{Lawrie84}. For
$m<m_c$, the random FP {\bf R} is stable and the expression for
$z_{\perp}$ and anisotropy function $\zeta_{\alpha}$ for $m\neq 1$
read \cite{Lawrie84}:
\begin{widetext}
\begin{eqnarray}\nonumber
z_{\perp}&=&2-\frac{m+2}{4(m-1)}\varepsilon+
\frac{(m+8)}{8(m-1)}{\tilde \varepsilon}+\{-4(m+2)[5m^2+42m+112-
192(m-1)\ln(4/3)]\varepsilon^2 \\
&-&4(m+2)[27m^2-264m-240+576(m-1)\ln(4/3)]\varepsilon{\tilde \varepsilon}\nonumber\\
&+&[59m^3-528m^2-2928m-
896+1728(m+2)(m-1)\ln(4/3)]{\tilde \varepsilon}^2\}(1024(m-1)^3)^{-1};\label{epp}\\
\zeta_{\alpha}&=&\frac{(m+2)}{4(m-1)}\varepsilon+\frac{m+8}{8(m-1)}{\tilde \varepsilon}\nonumber\\
&+&[-4(m+2)(5m^2+10m+144)\varepsilon^2-4(m+2)(27m^2-168m-336)\varepsilon{\tilde \varepsilon}\nonumber\\
&+&(59m^3-240m^2-2640m-1472){\tilde \varepsilon}^2][1024(m-1)^3]^{-1}, \qquad {\tilde \varepsilon}=\varepsilon+\varepsilon_d.\label{epp1}
\end{eqnarray}
\end{widetext}
 Whereas for $m>m_c$,  the pure
FP {\bf P} is stable. Therefore the critical behavior is
isotropic and the dynamic exponent $z$ reads:
\begin{equation}
z=2+ 0.363\,{\frac { \left( m+ 2 \right) {\varepsilon}^{2}}{
 \left( m+ 8 \right) ^{2}}}. \label{pur}
\end{equation}

As it is known \cite{sqrt}, due to degeneracy of the
$\beta$-functions of the weakly diluted Ising model with
point-like defects, the usual $\varepsilon$-expansion for critical
exponents turns into the $\sqrt{\varepsilon}$-expansion. The same
holds for the extended defects: indeed, the case   $m=1$ is to be
analyzed separately as far as  expressions (\ref{epp}),
(\ref{epp1}) contain poles at $m=1$, and one arrives
\cite{Lawrie84} at the $\sqrt{\varepsilon}$-expansion for the
critical exponents. In particular, the corresponding expressions
for $z$ can be given only to the lowest non-trivial order.
Moreover, the $\sqrt{\varepsilon}$-expansion does not allow for a
reliable numerical estimate \cite{Folk00}. Therefore, for the
disordered Ising model, the fixed dimension scheme considered in
the subsection \ref{IVA} remains the only way to get numerical
estimates.

To get the numerical estimate for $m_c$ in the frames of
$\varepsilon,\varepsilon_d$-expansion we substitute into the
modified Harris criterion the five-loop $\varepsilon$ - expansion
for the correlation length critical exponent $\nu_{p}$ for pure
$m$-component model, given in \cite{Kleinert91} and obtain the
following expansion:
\begin{widetext}
\begin{eqnarray}
m_c&=&\left( - 4+ 8{\frac {{\tilde \varepsilon}}{\varepsilon}}- 2.50000{\frac
{{{\tilde \varepsilon} }^{2}}{\varepsilon}}- 1.500000{\tilde \varepsilon}- 2.448919{\frac
{{{\tilde \varepsilon}}^{3}}{{\varepsilon}}}+ 3.014557
{{\tilde \varepsilon}}^{2}+ 4.141561\varepsilon{\tilde \varepsilon}-
 1.682130{\frac {{{\tilde \varepsilon}}^{4}}{\varepsilon}}\right.-\nonumber\\
&-&14.12940\,{\varepsilon}^{2}{\tilde \varepsilon}-
 0.5736055\,\varepsilon\,{{\tilde \varepsilon}}^{2}+ 7.657623\,{{\tilde \varepsilon}}^{3}+ 55.57104\,{{
\varepsilon}}^{3}{\tilde \varepsilon}- 16.25104\,{\varepsilon}^{2}{{\tilde \varepsilon}}^{2}-
37.62878\,{ \varepsilon}\,{{\tilde \varepsilon}}^{3}+\nonumber\\ &+& \left.
22.53257\,{{\tilde \varepsilon}}^{4}- 3.345417\,{\frac {{{\tilde \varepsilon}}^{ 5}}{\varepsilon}}
\right) \left(  2- {\frac {{\tilde \varepsilon}}{\varepsilon}}
 \right) ^{-1}, \qquad {\tilde \varepsilon}=\varepsilon+\varepsilon_d.\label{mcryt}
\end{eqnarray}
\end{widetext}
Putting here ${\tilde \varepsilon}=\varepsilon$ (i.e $\varepsilon_d=0$) one
recovers the $\varepsilon$-expansion for $m_c$ of the  model with
point-like uncorrelated defects \cite{Dudka01}.

To estimate $m_c$ numerically for different fixed values of
$d,\varepsilon_d$, one should apply a resummation. As it is known
\cite{Dudka01,Dudka04} already  simple Pad\'e-analysis gives
convergent results for marginal dimensions. In the Pad\'e analysis
\cite{Baker81}, it is known that the best convergence is achieved
along the main diagonal of a Pad\'e table, therefore we make use
of the diagonal [2/2] Pad\'e-approximant. In the three-dimensional
case ($\varepsilon=1$) it gives: $m_c(\varepsilon_d=0)=1.92$
(which is in a good agreement with the known six-loop results for
the point-like disorder $m_c=1.942\pm0.026$ \cite{Bervillier86}
and $m_c=1.912\pm 0.004$ \cite{Dudka01}),
$m_c(\varepsilon_d=0.1)=2.48$, $m_c(\varepsilon_d=0.2)=3.10$,
$m_c(\varepsilon_d=0.3)=3.77$, $m_c(\varepsilon_d=0.4)=4.54$. The
results obtained are plotted by the dotted line in Fig.
\ref{fig1}.

To estimate dynamical critical exponents in the FP {\bf R} we
applied the Chisholm-Borel resummation technique  to the functions
(\ref{epp}), (\ref{epp1}) treating them as series in two variables
$\varepsilon,\varepsilon_d$ and using the Chisholm approximant. In
the corresponding two-loop approximation the dynamical critical
exponents in the pure FP {\bf P} is given by Eq. (\ref{pur}) and
the expression is too short to be resummed. Therefore it has been
estimated at $\varepsilon=1$ and different $m$ by a direct
substitution. The results are presented in Table \ref{tab4}. They
qualitatively confirm our conclusions made on the basis of fixed
dimension technique in subsection \ref{IVA} (see Table
\ref{tab3}). The discrepancy between the data of the two
techniques may serve also as an estimate for accuracy of the
perturbation schemes applied.

Let us note, that currently the three-loop series for the exponent
$z$ of the pure $m$-vector magnet model A dynamics is available.
The expression found in \cite{Antonov84} reads:
\begin{equation} \label{ant}
z=2+0.726(1-\varepsilon\cdot0.1885)\eta.
\end{equation}
Substituting into (\ref{ant}) the  $\varepsilon$-expansion for
$\eta$ \cite{Kleinert91}, one gets:
\begin{widetext}
\begin{equation}
z=2+ 0.363\,{\frac { \left( m+ 2 \right) {\varepsilon}^{2}}{
 \left( m+ 8 \right) ^{2}}}+ \left(  0.09075\,{\frac { \left(
m+ 2 \right)  \left( - m^{2}+ 56\,m+ 272 \right) }{
 \left( m+ 8 \right) ^{4}}}- 0.363\,{\frac { 0.1885\,m+
  0.3770}{ \left( m+ 8 \right) ^{2}}} \right) {\varepsilon}^{3}.\label{3la}
\end{equation}
\end{widetext}
Numerical estimates of this series based on the Pad\'e-Borel
resummation are given in the Table \ref{tab1} for $m=1$ and compared with
values found by other approaches.

\begin{table}[h!]
\caption{\label{tab4} The dynamical critical exponents of
$m$-com\-po\-nent model with extended impurities, obtained on the
base of resumming the double $\varepsilon,\varepsilon_d$-expansion
(Eq. (\ref{epp})).}
\begin{center}
\begin{tabular}{c c  c  c  c c  c}
\hline  \hline &   & $m=2$ &  &$m=3$&  & $m=4$\\ \hline
$\varepsilon_d$ & $z_{\perp}$  &  $z_{||} $  & $z_{\perp}$ &
$z_{||} $  & $z_{\perp}$  & $z_{||}$ \\ 0 & 2.0145  &  - &  2.0150
& - & 2.0151 & -
\\ 0.1 &  2.021 & 2.0147 &   2.0150 &  2.0150 & 2.0151 & 2.0151\\
 0.2 &  2.046 &  2.0150 &  2.0150 &
2.0150 &  2.0151 & 2.0151\\
 0.3 &  2.067 &  2.0151 &  2.0511  &
2.020& 2.0151 &  2.0151\\ 0.4 & 2.175 & 2.094 &  2.088&  2.050 &
  2.024 &1.984 \\
 0.5 & 2.209 & 2.110 &   2.129 & 2.081 &
    2.064 & 2.020 \\
 0.6 & 2.238 & 2.122 &   2.167 & 2.101 &
 2.114 & 2.054  \\
 0.7 & 2.270 &  2.138 &  2.210 &  2.120 &
  2.158 & 2.085 \\ 0.8 & 2.287 & 2.146 &  2.247 &  2.140 &
2.195 & 2.113   \\
 0.9 & 2.315 &  2.157 &  2.282 & 2.153 &
 2.244 & 2.126 \\ 1.0 & 2.333 & 2.171 &  2.317 &  2.169 &
2.279 & 2.141 \\
 1.1 &  2.354 & 2.175 &  2.346 &  2.176 &
2.314&  2.164
\\ \hline\hline
\end{tabular}\end{center}
\end{table}

\section{Effective critical behavior}
\label{V}
 The previous discussion concerns the asymptotic critical
dynamics of the model (\ref{model}). Here we discuss the effective
critical behavior which is observed in approaching the
critical point $T_c$ \cite{effective}. This behavior is
characterized by effective critical exponents governing scaling
laws when $T_c$ still is not reached. The calculation of effective
critical exponents for models with extended impurities was
not considered so far in dynamics as well as in the statics. However these
exponents are mainly observed in numerical and real experiments, which
may be outside the asymptotic region.

The effective critical exponents are defined as logarithmic
derivatives of the appropriate quantities of interest with respect
to the reduced temperature $t=|T-T_c|/T_c$ \cite{effective}. For
instance, effective critical exponent for perpendicular correlation length is defined  as:
\begin{equation}\label{def}
\nu_{\perp}^{\rm eff}(t)=-\frac{{\rm d}\ln \xi_{\perp}(t)}{{\rm d}\ln t}.
\end{equation}
In the limit $T\to T_c$ the effective exponent coincides with the
asymptotic one $\nu_{\perp}^{\rm eff}=\nu_{\perp}$.
 Within
the field-theoretical RG approach effective exponents are calculated
using the values of couplings (solution of flow equation)
depending on the flow parameter $\ell$. For instance, effective
exponents  for correlation length and relaxation time perpendicular
to the extended impurities direction are defined
as:
\begin{eqnarray}\label{eff}
&&1/\nu^{\rm eff}_{\bot}(\ell)=2- {\bar
\gamma}_{\phi^2}(u(\ell),v(\ell))-\nonumber\\
&&-\gamma_{\phi}(u(\ell),v(\ell))+\dots;
\\
&&z^{\rm
eff}_{\bot}(\ell)=2+\zeta(u(\ell),v(\ell))+\dots\qquad\label{eff2}.
\end{eqnarray}
The flow parameter $\ell$ is related to the temperature distance to the
critical point via the inverse correlation length.
Therefore the dependence of the effective exponents on the
flow parameter corresponds to a dependence on the reduced
temperature \cite{remark}. In (\ref{eff}) and (\ref{eff2}) the  parts denoted by
dots  come from the change of the amplitude part of the
perpendicular correlation length and critical
relaxation time correspondingly. In the subsequent calculations we
will neglect this part, since the contribution of the amplitude
function to the crossover seems to be small \cite{note2}.

Solving the flow equation for different initial conditions we can
obtain  different flows in the space of couplings. We choose these
condition near the Gaussian FP ({\bf G}), expecting that couplings
are small in the background region. In flow equations (\ref{13a})
we use the $\beta$-function (\ref{betau}), (\ref{betav}) at fixed
$d$  resummed by the Chisholm-Borel method. This allows us to
investigate also the case of Ising systems $(m=1)$, for which the
$\beta$-function are degenerated on the one-loop level and the
$\sqrt\varepsilon$-expansion does not allow for numerical
estimates \cite{Folk00}.

\begin{figure}
{\includegraphics[width=0.4\textwidth]{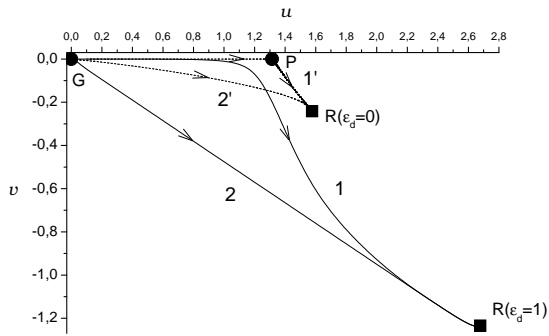}}
\caption{\label{flowex} Flows for the case $m=1$. Dashed lines
correspond to the case of uncorrelated point defects,
$\varepsilon_d=0$, while solid lines correspond to system with
extended lines of defects, $\varepsilon_d=1$. The initial values
are the same for curves 1 and $1'$ (small ratio $\Delta/u$) and for
2 and $2'$ (large ratio $v/u$). }
\end{figure}

The RG flows at $m=1$ for the two most interesting cases
$\varepsilon_d=0$ and $\varepsilon_d=1$ are shown in the
Fig.~\ref{flowex} (dashed
lines for $\varepsilon_d=0$ and solid lines for $\varepsilon_d=1$).
The first case corresponds to the point
defects, whereas the second one corresponds to the lines of
defects. We consider the same initial condition for both cases
with a small (flows 1 and $1'$)  and with a large (flows 2 and $2'$)
value of the ratio $v/u$. The ratio $v/u$ defines the degree of disorder,
thus we can observe difference in the behavior of systems with low
and high dilution.

\begin{figure}
{\includegraphics[width=0.4\textwidth]{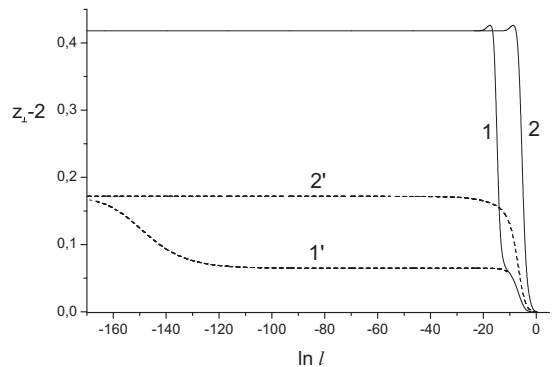}}
\caption{\label{zex} Behavior of the dynamical effective exponent
$z^{\rm eff}_{\bot}$ for $m=1$ in the case of  $\varepsilon_d=0$
(dashed lines) and $\varepsilon_d=1$ (solid lines). Curves
correspond to flows of Fig.\protect\ref{flowex}.}
\end{figure}

\begin{figure}
{\includegraphics[width=0.4\textwidth]{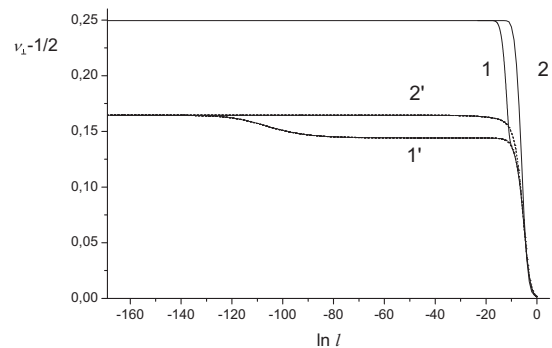}}
\caption{\label{nex} Behavior of  effective critical exponent of
perpendicular correlation length $\nu^{\rm eff}_{\bot}$ for $m=1$
in the case of $\varepsilon_d=0$ (dashed lines) and
$\varepsilon_d=1$ (solid lines). Curves correspond to flows of
Fig.\protect\ref{flowex}.}
\end{figure}

Using the flows given in the Fig.~\ref{flowex} we can obtain static
and dynamic effective exponents. Below, we choose to display
 the exponents $\nu^{\rm eff}_{\bot}$ and $z^{\rm eff}_{\bot}$.
Due to Eq. (\ref{zpar}) the parallel effective exponents $\nu^{\rm
eff}_{||}$ and $z^{\rm eff}_{||}$ have qualitatively similar
behavior. The deviations of effective exponents from the
mean-field values $z^{\rm eff}_{\bot}-2$  and $\nu^{\rm
eff}_{\bot}-1/2$ corresponding to flows of Fig.~\ref{flowex} are
given in the Fig.~\ref{zex} and Fig.~\ref{nex} respectively. First
we compare the behavior of the effective exponents of the uniaxial
$m=1$ magnets at $\varepsilon_d=0$ and $\varepsilon_d=1$. As it
can be seen from the Figs.~\ref{zex} and \ref{nex}, even a weak
dilution by lines of defects leads to a faster crossover to the new
universality class (curves 1 in Figs. ~\ref{zex} and \ref{nex}) and
the effective exponents are not influenced by the pure FP {\bf P}.
Such a behavior qualitatively differs from that at dilution by
point-like impurities, where effective exponents corresponding
to the pure FP {\bf P} are observed in a relatively wide region
(curves 1` in both figures). Note that the dynamical critical
exponent $z^{\rm eff}_{\bot}$ at $\varepsilon_d=1$ approaches its
asymptotics from above. Such a nonmonotonic behavior is typical in
statics for system with point-like defects \cite{Folk03,Folk00}.
However we do not observe nonmonotonic behavior for $\nu^{\rm
eff}_{\bot}$. It may be a feature of static effective critical
exponents for system with extended impurities or consequence of
our approximation.

\begin{figure}
{\includegraphics[width=0.4\textwidth]{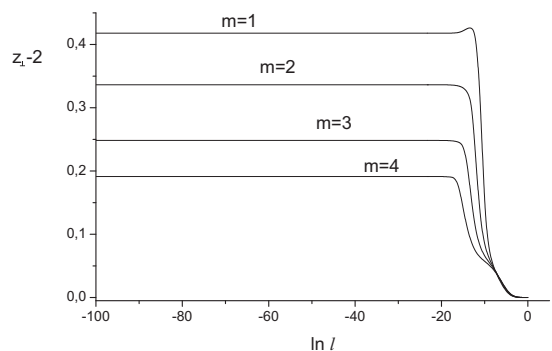}}
\caption{\label{zm} Behavior of dynamical effective exponent
$z^{\rm eff}_{\bot}$ for system with extended impurities with
$\varepsilon_d=1$ and different $m$. Initial conditions are the
same in all cases (small ratio $v/u$).}
\end{figure}

\begin{figure}
{\includegraphics[width=0.4\textwidth]{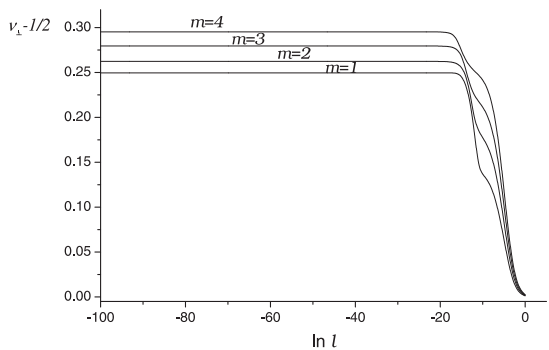}}
\caption{\label{nm} Behavior of  effective critical exponent of
perpendicular correlation lenght $\nu^{\rm eff}_{\bot}$ for system
with extended impurities with $\varepsilon_d=1$ and different $m$.
Initial conditions are the same in all cases (small ratio
$v/u$).}
\end{figure}

We also consider static and dynamic effective behavior
for different values of order parameter dimension. Fig.~\ref{zm}
and Fig.~\ref{nm} present the $z^{\rm eff}_{\bot}-2$ and
$\nu^{\rm eff}_{\bot}-1/2$  correspondingly obtained for the same
initial conditions and different values $m=1,\,2,\,3,\,4$. The
nonmonotonic character of dependence of exponents on the logarithm
of flow parameter is observed  only for $z^{\rm eff}_{\bot}$ at
$m=1$.

\section{Conclusions}
\label{VI}

The present study is dedicated to providing the reliable numerical
estimates for the dynamical critical exponents of magnetic systems
with $\varepsilon_d$-dimensional non-magnetic impurities of
parallel orientation. The presence of such impurities leads to
anisotropy in the systems; thus, two correlation lengths, parallel
and perpendicular to the extended defects exist, which diverge in
the vicinity of critical temperature with different exponents. The
dynamic critical behavior is modified as well, so that two times
of critical relaxation in directions parallel and transverse to
extended impurities appear. The former results obtained for such
systems are based on the double expansion in parameters
$\varepsilon,\varepsilon_d$ and thus are rather of a qualitative
character.

An example of related systems is given by magnets with extended
randomly oriented defects: these can be described by the model of
Weinrib and Halperin \cite{Weinrib83}. There, the distribution of
defects is characterized by the pair correlation function, falling
off with distance for large separations according to a power law.
The influence of such correlated defects on criticality was a
subject of several theoretical studies \cite{Weinrib83,lr} and
predicted new universal critical behavior found recently its
confirmation in the MC simulations \cite{mcext}.

 We applied the resummation to the RG
functions of the model, obtained in the minimal subtraction scheme
\cite{Boyanovsky82,Lawrie84}, treating them directly for fixed
$d=3$ and fixed parameter $\varepsilon_d$.  The case
$\varepsilon_d=0$ describes point-like quenched disorder and
reproduces well-known results. For $\varepsilon_d>0$ it was found,
that the relation $z_{\perp}>z_{||}$ holds for every
$\varepsilon_d$ and $m$, which is connected to the fact, that the
extended defects cut interacting paths of spins perpendicular to
the extended-defect direction, so in the parallel direction the
fluctuations are stronger.

The data of Tables \ref{tab1}, \ref{tab4} give numerical values of
the dynamical critical exponents  for the model A dynamics of the
$m$-vector magnets with extended $\varepsilon_d$-dimensional
defects.  Together with possible scenarios of the effective
critical behavior discussed in the Section \ref{V} it should
facilitate experimental studies of the influence of extended
defects on criticality. As we already noted, the point defects
change the universality class at $d=3$ only for the Ising magnets
($m=1$). Values of the critical exponent $z$ of the pure and
the diluted (by point-like impurities) Ising model are given in Tables
\ref{tab2}, \ref{tab3}, respectively. Comparing the data in these
tables one sees an increase of the exponent for the diluted model with respect to the pure one.
 This corresponds to a
stronger divergency of the relaxation time (increase of the
critical slowing-down effects). Further, comparing Tables
\ref{tab3} and \ref{tab1} one sees that extended impurities make
this effect more pronounced, leading to further increase of both
exponents $z_{\perp},z_{||}$. However this increase (of the order
of 8 \% if one compares $z$-exponents for point and line defects)
is not dramatic to make MC simulations impossible due to enhanced
critical slowing down. On the other hand, the same order of
difference in critical exponents values holds also for the static
exponents \cite{Blavatska02,Blavatska03}. Therefore, we expect
that the predicted change in the critical behavior of three
dimensional magnets with extended impurities is within current
experimental accuracy. Although it is more difficult to extract
dynamical critical exponents from simulations to the same
accuracy, it would be worthwhile to look for them (i) in the cases
of different defect dimensionality and (ii) in the  non-asymptotic
regime.

\section*{Acknowledgements}
We thank the Austrian Fonds zur F\"orderung der wissenschaftlichen
Forschung, project No.16574-PHY which supported in part this
research. V.B. thanks the W. Macke Stiftung for enabling her
research stay in Linz.


\begin{thebibliography}{77}
\bibitem{note}
Our Lagrangian (\ref{model}) differs from form obtained in Ref.
\protect\cite{Lawrie84} since we introduce the auxialiary field
$i\tilde \phi$ instead of $\phi$ of Ref.
\protect\cite{Lawrie84}.
\bibitem{remark}
In general this dependence is not given by a power law but is more complicated
and would have to be calculated in the non-asymptotic theory.
\end{thebibliography}
\end{document}